# Matrix Access structure Policy used in Attribute-Based Proxy Re-encryption

Keying Li

Faculty of science, Xidian University, Xi'an 710071, China

**Abstract**

Proxy re-encryption (PRE) allows a semi-trusted proxy to convert a ciphertext originally intended for Alice into an encryption of the same message intended for Bob. Song Luo, Jianbin Hu, and Zhong Chen presented a novel ciphertext policy attribute-based proxy re-encryption (CP-AB-PRE) scheme. The ciphertext policy realized in their scheme is AND-gates policy supporting multi-value attributes, negative attributes and wildcards. We propose a new access policies based on LSSS matrix access structures. Our scheme still have the properties of both PRE and CP-AB-PRE, such as unidirectionality, non-interactivity, multi-use, allows the encryptor to decide whether the ciphertext can be re-encrypted and allows the proxy to add access policy. Furthermore, our scheme can be modified to outsource the policy of $W_2$.

*Keywords: Proxy Re-encryption, Attribute-Based Encryption, Ciphertext Policy, matrix access structures*

## 1. Introduction

After Boneh and Franklin [1] proposed a practical identity-base encryption (IBE) scheme, Green and Ateniese [2] proposed the first identity-based PRE (IB-PRE).A proxy re-encryption (PRE) scheme allows a proxy to translate a ciphertext encrypted under Alice's public key into one that can be decrypted by Bob's secret key. The proxy may be an untrusted third party. PRE was used in many scenarios. Imagine that one day you are on a business trip and is inconvenient to access your email. You would wish to have the mail server forward your encrypted email messages to your secretary Bob, who can then read the message using his own or new private key. Once Alice comes back, the proxy was asked to stop transferring the emails. Difference with the traditional proxy decryption scheme, PRE does not need users to store any additional decryption key, in other words, any decryption would be finished using only his own secret keys. But in our scheme, the User2 need some Auxiliary key.

Attribute-based encryption (ABE) is a generalization of IBE. The data provider can express how he wants to share data in the encryption algorithm itself. Goyal, Pandey, Sahai, and Waters [3] further clarified the concept of Attribute-Based Encryption. There are two kind of ABE schemes, key policy ABE (KP-ABE) and ciphertext policy ABE (CP-ABE) schemes. In KP-ABE schemes, ciphertexts are associated with sets of attributes and users' secret keys are associated with access policies. In CP-ABE schemes, the situation is reversed. That is, each ciphertext is associated with an access policies. CP-ABE have three kind of access structure. Cheung and Newport [4] use AND-gates as the access strategy, for the first time proved the security of CP-ABE mechanism under DBDH assumption. Bethencourt, Sahai, and Waters [5] use the tree structure to realize Fine-Grained access control strategy. Waters [6] use LSSS access structure $(M, \rho)$ under DPBDHE (decisional Parallel Bilinear Diffie-Hellman Exponent) hypothesis. We can implement proxy re-encryption in ABE schemes, as ABE is the development of IBE. But it is not a trivial work to apply proxy re-encryption technique into attribute based system. Song Luo, Jianbin Hu [39] proposed a novel ciphertext policy attribute-based proxy re-encryption (CP-AB-PRE) scheme. The ciphertext policy realized in their scheme is AND-gates policy supporting multi-value attributes, negative attributes and wildcards.

**Our Contributions** We present a ciphertext policy attribute-based proxy re-encryption (CP-AB-PRE) scheme using matrix access structure. The ciphertext policy realized in our scheme is matrix access policy, which also supporting multi-value attributes, negative attributes and wildcards. It is more convents to control the attributes of User2, and realize the policy of W2 more efficiently. Our scheme inherits the following properties of PRE mentioned in [2, 7]:

– **Unidirectionality**. Ueser1 can delegate decryption rights to User2 without permitting her to decrypt User2's ciphertext.

– **Non-Interactivity.** User1 can compute re-encryption keys without the participation of User2 or the private key generator (PKG).

– **Multi-Use.** The proxy can re-encrypt a ciphertext multiple times, e.g. re-encrypt from User1 to User2, and then re-encrypt the result from User2 to User3.In this process, the computation would increase, but not exponent increasing.

**Our scheme has the other three properties:**

– **Secret Key Security** [7]. A valid proxy designated by User1, other users who are able to decrypt User1's ciphertext with the help from the proxy cannot collude to obtain User1's secret key.

– **Re-encryption Control.** User1 can decide whether the ciphertext can be re-encrypted.

– **Extra Access Control.** When the proxy re-encrypts the ciphertext, he can add extra access policy to the ciphertext.

– **Re-outsourcing.** In our scheme, the User1 can finish the process of proxy. If we omit the proxy, not only User1but also User2 would increase the amount of computation. We will use the third party. We can also outsource the policy of $W_2$.

**Related Work** Sahai and Waters [8] first proposed Attribute-based encryption and later clarified in [3]. The first CP-ABE scheme was put forward by Bethencourt, Sahai, and Waters [5]. Their scheme allows the ciphertext policies to be very expressive, but the security proof is in the generic group model. Cheung and Newport [4] raised a provably secure CP-ABE scheme which is proved to be secure under the standard model. Further on, their scheme supports AND-Gates policies which deals with negative attributes explicitly and uses wildcards in the ciphertext policies. Goyal et al. [9] put forward a bounded ciphertext policy attribute-based encryption in which a general transformation method was proposed to transform a KP-ABE system into a CP-ABE one by using "universal" access tree. However, the parameters of ciphertext and private key sizes will grow up in the worst case. The first secure CP-ABE scheme was presented by Waters [6], which supported general access formulas. Lewko et al. [10] present a fully secure CP-ABE scheme by using the dual system encryption techniques [11, 12].

There are also many other ABE schemes. Multiple authorities were introduced in [13] and [14]. K.Emura et al. [15] introduced a novel scheme using AND-Gates policy. In their scheme, it has constant ciphertext length. Hiding access structure in attribute-based encryption is also a problem. T.Nishide et al. [16] gave a method to solve. Attribute-based encryption was enhanced by R.Bobba et al. [17] with attribute-sets which allow same attributes in different sets. N.Attrapadung et al. [18] proposed dual-policy attributebased encryption which allows key-policy and ciphertext-policy act on encrypted data simultaneously. Matthew Green,Susan Hohenberger and Brent Waters [19] proposed the concept of outsourcing, which had relationship with the proxy(third party). It can also be regard as the expansion of the PRE.

Recently, predicate encryption was proposed by Katz, Sahai, and Waters [20] and furthered by T.Okamoto et al. [21].

Mambo and Okamoto [22] first introduced the notion of PRE. Later Blaze et al. [23] presented the first concrete bidirectional PRE scheme which allows the key holder to announce the proxy function and have it applied by untrusted third parties without further involvement by the original key holder. These schemes all had multi-use property. The first unidirectional and single-use proxy re-encryption scheme was presented by Ateniese et al. [7]. Boneh, Goh and Matsuo [24] described a hybrid proxy re-encryption system based on the ElGamal-type PKE system [25] andBoneh-Boyen's identity-based encryption system [26]. In 2007, Green and Ateniese [2] provided identity-based PRE but their schemes are secure in the random oracle model. Chu et al. [34] proposed new identity-based proxy re-encryption schemes in the standard model. Matsuo [27] proposed new proxy re-encryption system for identity-based encryption, but the scheme needs a re-encryption key generator (RKG) to generate re-encryption keys. Libert and Vergnaud [28] proposed a traceable proxy re-encryption system, in which a proxy that leaks its re-encryption key can be identified by the delegator. After the present of ABE, Guo et al. [29] proposed the first attribute-based proxy re-encryption scheme, their scheme is based on key policy and bidirectional. Liang et al. [30] proposed the first ciphertext policy attribute-based proxy re-encryption scheme which has the above properties except re-encryption control.

**Organization** The paper is organized as follows. We give necessary background information and assumptions in Section 2. We present our scheme and secure model, then construct and give a proof of security in Section 3. Discuss a number of extensions of the proposed scheme in Section 4.

## 2. Preliminaries

### 2.1 Bilinear Maps

Let $\mathbb{G}$ and $\mathbb{G}_T$ be two multiplicative cyclic groups of prime order $p$. Let $g$ be a generator of $\mathbb{G}$ and e: $\mathbb{G} \times \mathbb{G} \to$

$\mathbb{G}_T$ be a bilinear map with the properties:

1. Bilinearity: for all $u, v \in \mathbb{G}$ an $a, b \in \mathbb{Z}_p$, we have $e(u^a, v^b) = e(u,v)^{ab}$.

2. Non-degeneracy: $e(g,g) \neq 1$.

We say that $\mathbb{G}$ is a bilinear group if the group operation in $\mathbb{G}$ and the bilinear map $e: \mathbb{G} \times \mathbb{G} \to \mathbb{G}_T$ are both efficiently computable.

**2.2 Access Structure**

**Definition2** (*Access Structure* [31]) Let $\{P_1, P_2, \cdots, P_n\}$ be a set of parties. A collection $A \subseteq 2^{\{P_1,P_2,\cdots,P_n\}}$ is monotone if $\forall B, C$: if $B \in A$ and $B \subseteq C$ then $C \in A$. An access structure (respectively, monotone access structure) is a collection (resp., monotone collection) $A$ of non-empty subsets of $\{P_1, P_2, \cdots, P_n\}$, i.e., $A \subseteq 2^{\{P_1,P_2,\cdots,P_n\}}$. The sets in $A$ are called the authorized sets, and the sets not in $A$ are called the unauthorized sets.

In our context, the role of the parties is taken by the attributes. Thus, the access structure A will contain the authorized sets of attributes. We restrict our attention to monotone access structures. However, it is also possible to (inefficiently) realize general access structures using our techniques by defining the "not" of an attribute as a separate attribute altogether. From now on, unless stated otherwise, by an access structure we mean a monotone access structure.

**2.3 LSSS and Monotone Span Programs[6]:**

In a linear secret-sharing scheme [31], realizing an access structure A, a third party called the dealer holds a secret *y* and distributes the shares of *y* to parties such that *y* can be reconstructed by a linear combination of the shares of any authorized set. Further, an unauthorized set has no information about the secret *y*.

There is a close relation between LSSS and a linear algebraic model of computation called monotone span programs (MSP) [32]. It has been shown that the existence of an efficient LSSS for some access structure is equivalent to the existence of a small monotone span program for the characteristic function of that access structure [31, 32]

**2.4 Decisional Parallel Bilinear Diffie-Hellman Exponent Assumption [6]**

We define the decisional q-parallel Bilinear Diffie-Hellman Exponent problem as follows. Choose a group $\mathbb{G}$ of prime order p according to the security parameter. Let $a, s, b_1, \cdots, b_q \in \mathbb{Z}_p$ be chosen at random and g be a generator of $\mathbb{G}$. If an adversary is given

$$\vec{y} = g, g^s, g^a, \cdots, g^{(a^q)}, , g^{a^{(q+2)}}, \cdots, g^{a^{(2q)}}$$

$\forall\ 1 \leq j \leq q$

$$g^{s \cdot b_j}, g^{a/b_j}, \cdots, g^{(a^q/b_j)}, , g^{(a^{q+2}/b_j)}, \cdots, g^{(a^{2q}/b_j)}$$

$\forall\ 1 \leq j, k \leq q, k \neq j \quad g^{a \cdot s \cdot b_k/b_j}, \cdots, g^{(a^q \cdot s \cdot b_k/b_j)}$

it must remain hard to distinguish $e(g,g)^{a^{q+1}s} \in \mathbb{G}_T$ from a random element in $\mathbb{G}_T$. An algorithm $\mathcal{B}$ that outputs $z \in \{0,1\}$ has advantage $\epsilon$ in solving decisional q-parallel BDHE in $\mathbb{G}$ if

$$|\Pr[\mathcal{B}(\vec{y}, T = e(g,g)^{a^{q+1}s}) = 0] - \Pr[\mathcal{B}(\vec{y}, T = R) = 0]| \geq \epsilon$$

**Definition 2.1** We say that the (decision) q parallel-BDHE assumption holds if no polytime algorithm has a non-negligible advantage in solving the decisional q-parallel BDHE problem.

## 3. CP Attribute-Based Proxy Re-encryption

### 3.1 Algorithms of CP-AB-PRE

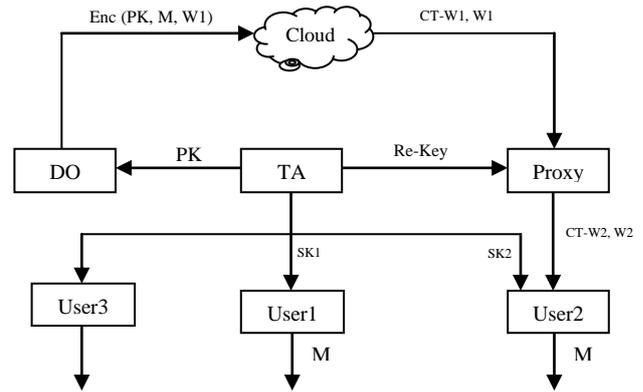

Fig.1 encryption system

A CP-AB-PRE scheme consists of the following six algorithms:

**Setup**, **KeyGen**, **Encrypt**, **RKGen**, **Reencrypt**, and **Decrypt**.

**Setup** $(1^\kappa)$. This algorithm takes the security parameter $\kappa$ as input and generates a public key *PK*, a master secret key *MSK*.

**KeyGen**(*MSK*, $L_1$). This algorithm takes *MK* and a set of attributes $L_1$ as input and generates a secret key $SK_{L_1}$ associated with $L_1$.

**Encrypt** ($PK, M, W_1$). This algorithm takes *PK*, a message *M*, and an access policy $W_1$ as input, and generates a ciphertext $CT_{W_1}$.

**RKGen**($SK_{L_1}, L_2$). This algorithm takes a secret key $SK_{L_1}$ and a set of attributes $L_2$ as input and generates a re-encryption key $RK_{L_1 \to L_2}$.

**Reencrypt** ($RK_{L_1 \to L_2}, CT_{W_1}, W_2$). This algorithm takes a re-encryption key $RK_{L_1 \to L_2}$, and a ciphertext $CT_{W_1}$ as input, first checks if the attribute list in $L_1$ satisfies the access policy of $CT_{W_1}$, that is, $L_1| = W_1$. Then, if check passes, it generates a re-encrypted ciphertext $CT_{W'_1}$; otherwise, it returns ⊥. In addition, it generates the part ciphertext $CT_{W_2}$ of the policy $W_2$

**Decrypt** $(CT_{W_1}, SK_{L_1}; CT_{W'}, CT_{W_2}, SK_{L_2})$ This algorithm takes $CT_{W'}, SK_{L_2}, CT_{W_2}$ associated with $L_2$ as input and returns the message $M$ if the attribute list $L_2$ satisfies the access policy $W_2$ specified for $CT_{W_2}$, that is, $L_2| = W_2$. If $L_2| \neq W_2$, it would returns ⊥ with overwhelming probability.

### 3.2 Security Model

We describe the security model called Selective-Policy Model for our CP-ABPRE scheme. Based on [30], we use the following security game. A CP-AB-PRE scheme is selective-policy chosen plaintext secure if no probabilistic polynomial time adversary has non-negligible advantage in the following Selective-Policy Game.

**Selective-Policy Game for CP-AB-PRE**

**Init:** The adversary $\mathcal{A}$ commits to the challenge ciphertext policy $W_1^*$.

**Setup:** The challenger runs the **Setup** algorithm and gives PK to $\mathcal{A}$.

**Phase 1:** $\mathcal{A}$ makes the following queries.

− Extract($L_1^*$): $\mathcal{A}$ submits an attribute list $L_1^*$ for a KeyGen query where $L_1^*| \neq W_1^*$, the challenger gives the adversary the secret key $SK_{L_1^*}$.

− RKExtract($L_2^*, W_2^*$): $\mathcal{A}$ submits an attribute list $L_2^*$ for a RKGen query where $L_2^*| \neq W_2^*$, the challenger gives the adversary the re-encryption key $RK_{L_1^* \to L_2^*}$.

**Challenge**: $\mathcal{A}$ submits two equal-length messages $M_0, M_1$ to the challenger. The challenger flips a random coin b and passes the ciphertext Encrypt(PK, $M_b, W_2^*$) to the adversary.

**Phase 2**: Phase 1 is repeated.

**Guess**: $\mathcal{A}$ outputs a guess $b'$ of b.

The advantage of $\mathcal{A}$ in this game is defined as

$$\text{Adv}_\mathcal{A} = |\Pr[b' = b] - \frac{1}{2}|$$

In [13], Ateniese et al. defined another important security notion, named delegator secret security (or master key security), for unidirectional PRE. This security notion captures the intuition that, even if the dishonest proxy colludes with the User2, it is still impossible for them to derive the delegator's private key in full.

We give master key security game for attribute-based proxy re-encryption as follows. A CP-AB-PRE scheme has selective master key security if no probabilistic polynomial time adversary $\mathcal{A}$ has a non-negligible advantage in winning the following selective master key security game.

**Selective Secret Key Security Game**

**Init**: The adversary $\mathcal{A}$ commits to a challenge attribute list $L_2^*$.

**Setup**: The challenger runs the **ReKeyGen** algorithm and gives $RK_{L_1^* \to L_2^*}$ to $\mathcal{A}$.

**Queries**: $\mathcal{A}$ makes the following queries.

− Extract($L_1^*$): $\mathcal{A}$ submits an attribute list $L_1^*$ for a KeyGen query where $L_1^* \neq L_1$, the challenger gives the adversary the secret key $SK_{L_1^*}$.

− RKExtract($L_2^*, W_2^*$): $\mathcal{A}$ submits an attribute list $L_2^*$ for a RKGen query, the challenger gives the adversary the re-encryption key $RK_{L_1^* \to L_2^*}$.

**Output**: $\mathcal{A}$ outputs the secret key $SK_{L_1^*}$ for the attribute list $L_1^*$, then $\mathcal{A}$ succeeds.

The advantage of $\mathcal{A}$ in this game is defined as

$$\text{Adv}_\mathcal{A} = \Pr[\mathcal{A} \text{ succeeds}].$$

### 3.3 Proposed Scheme

Let $\mathbb{G}$ be a bilinear group of prime order p, and let g be a generator of $\mathbb{G}$. In addition, let $e : \mathbb{G} \times \mathbb{G} \to \mathbb{G}_1$ denote the bilinear map. Let $E : \mathbb{G} \to \mathbb{G}_1$ be an encoding between $\mathbb{G}$ and $\mathbb{G}_1$. A security parameter, $\kappa$, will determine the size of the groups. Let $\mathcal{U} = \{att_1, \cdots, att_n\}$ be a set of attributes; $S_i = \{v_{i,1}, \cdots, v_{i,n_i}\}$ be a set of possible values associated with $att_i$ and $n_i = |S_i|$; $L = [L_1, \cdots, L_n]$ be an attribute list for a user; and $W = [W_1, \cdots, W_n]$ be an access policy. [33]

Our six algorithms are as follows:

**Setup**($1^\kappa$) The setup algorithm takes as input the number of attributes in the system. It then chooses a group $\mathbb{G}$ of prime order p, a generator g and U random group elements $h_1, \cdots, h_u \in \mathbb{G}$ that are associated with the U attributes in the system. In addition, it chooses random exponents $\alpha$, $a \in \mathbb{Z}_p$.

The public key is published as

$$PK = g\,;\, e(g;g)^\alpha;\, g^a;\, h_1,\cdots, h_u$$

The authority sets $MSK = g^\alpha$ as the master secret key.

**KeyGen($MSK, L_1$)**. The key generation algorithm takes as input the master secret key and a set S of attributes. The algorithm first chooses a random $t \in \mathbb{Z}_p$. It creates the private key as

$$SK_{L_1}: K = g^\alpha g^{at} \quad L = g^t \quad \forall x \in L_1\; K_x = h_x^t$$

**Encrypt($PK, M, W_1$)** The encryption algorithm takes as input the public parameters PK and a message M to encrypt. In addition, it takes as input an LSSS access structure $W_1 = (M_1; \rho_1)$. The function associates rows of M to attributes.

Let $M_1$ be a $\ell \times n$ matrix. The algorithm first chooses a random vector $\vec{v} = (s, y_2, \cdots, y_n) \in \mathbb{Z}_p^n$. These values will be used to share the encryption exponent s. For i = 1 to $\ell$, it calculates $\lambda_i = \vec{v} \cdot M_{1i}$, where $M_{1i}$ is the vector corresponding to the i th row of $M_1$. In addition, the algorithm chooses random $r_1, \cdots, r_\ell \in \mathbb{Z}_p$.

The cipher text is published as $CT_{W_1}$:

$$C = Me(g;g)^{\alpha s};\, C' = g^s;\, g_2^s;$$
$$\left(C_1 = g^{a\lambda_1}h_{\rho_1(1)}^{-r_1}, D_1 = g^{r_1}\right), \cdots,$$
$$\left(C_\ell = g^{a\lambda_\ell}h_{\rho_1(\ell)}^{-r_\ell}, D_\ell = g^{r_\ell}\right)$$

along with a description of $(M_1; \rho_1)$.

**RKGen ($MSK, SK_{L_1}, L_2$)**. The Re-key generation algorithm takes as input the $MSK, SK_{L_1}, L_2$, choose random $d \in \mathbb{Z}_p$, and compute $g_2^d$, send it to User2, send $adt_1$, and the new private key $RK_{L \to L_2}$ to proxy.

$$K' = g^\alpha g^{at} g_2^{ad};\, L' = g^t g_2^d;\, \forall x \in L_1\; K_x' = h_x^t g_2^d$$

$$SK_{L_2}:\quad L'' = g^{t_1^{-1}} \;\forall x' \in L_2\; K_{x'}'' = h_{x'}^{t_1^{-1}}$$

**Reencrypt ($RK_{L \to L_2}, CT_{W_1}, W_2$)** The Re-encryption algorithm takes as input the public parameters $PK, RK_{L \to L_2}, CT_{W_1}$ to Re-encrypt. In addition, it takes as input an LSSS access structure $W_1 = (M_1; \rho_1), W_2 = (M_2; \rho_2)$. The function associates rows of $M_2$ to attributes.

Let $M_2$ be a $\ell \times n$ matrix. The algorithm first chooses a random vector $\vec{v}' = (adt_1, d_2, \cdots, d_n) \in \mathbb{Z}_p^n$. These values will be used to share the encryption exponent s. For i = 1 to $\ell$, it calculates $\lambda_i' = \vec{v}' \cdot M_{2i}$, where $M_{2i}$ is the vector corresponding to the ith row of $M_2$. In addition, the algorithm chooses random $r_1', \cdots, r_\ell' \in \mathbb{Z}_p$.

The Re-encryption algorithm then takes as input $L_1, L_2$. Suppose that $L_1$ satisfies the access structure and let $I_1 \subset \{1, 2, \cdots, \ell\}$ be defined as $I_1 = \{i: \rho_1(i) \in L_1\}$. Then, let $\{\omega_i \in \mathbb{Z}_p\}_{i \in I_1}$ be a set of constants such that if $\lambda_i$ are valid shares of any secret s according to $M_1$, then $\sum_{i \in I_0} \omega_i \lambda_i = s$.

The cipher text is published as $CT_{W_2}$:

$$\frac{e(C', K')}{\prod_{i \in I_1}(e(C_i, L')e(D_i, K_{\rho_1(i)}'))^{\omega_i}}$$

$$= \frac{e(g,g)^{\alpha s}}{\prod_{i \in I_1} e(h_{\rho_1(i)}, g_2)^{-r_i d \omega_i} e(g, g_2)^{r_i \omega_i d}}$$

$$\left(C_1' = g_2^{-s\lambda_1'}h_{\rho_2(1)}^{-r_1'}, D_1' = g^{r_1'}\right), \cdots,$$
$$\left(C_\ell' = g_2^{-s\lambda_\ell'}h_{\rho_2(\ell)}^{-r_\ell'}, D_\ell' = g^{r_\ell'}\right)$$

**Decrypt($CT_{W_1}, SK_{L_1}; CT_{W_2}, SK_{L_2}$)**. The decryption algorithm takes as input a cipher text $CT_{W_1}, CT_{W_2}$ for access structure $(M_1; \rho_1), (M_2; \rho_2)$ and a private key for a set $L_2$. Suppose that $L_2$ satisfies the access structure and let $I_2 \subset \{1, 2, \cdots, \ell\}$ be defined as $I_2 = \{i: \rho_2(i) \in L_2\}$ $I_2 = \{i: \rho_2(i) \in L_2\}$. Then, let $\{\omega_i' \in \mathbb{Z}_p\}_{i \in I_2}$ be a set of constants such that if $\lambda_i'$ are valid shares of any secret d, then $\sum_{i \in I_1} \omega_i' \lambda_i' = d$

The decryption algorithm first computes

$$\frac{e(g,g)^{\alpha s} \prod_{i \in I_2}\left(e(C_i', L'')e(D_i', K_{\rho_1(i)}'')\right)^{\omega_i'}}{\prod_{i \in I_1} e(h_{\rho_1(i)}, g_2)^{-r_i d \omega_i} e(g, g_2)^{r_i \omega_i d} \prod_{i \in I_1} e(C_i D_i, g_2^{-d})^{\omega_i}}$$

$$= \frac{e(g,g)^{\alpha s} \prod_{i \in I_2}\left(e\left(g_2^{-s\lambda_i'}h_{\rho_2(i)}^{-r_i'}, g^{t_1}\right) e\left(g^{r_i'}, h_{\rho_2(i)}^{t_1}\right)\right)^{\omega_i'}}{e(g, g_2)^{-ads}}$$

$$= e(g,g)^{\alpha s}$$

The decryption algorithm can then divides out this value from C and obtain the message M.

**3.4 Security proof**

**Theorem 1** If there is an adversary who breaks our scheme in the Selective-Policy model, a simulator can take the adversary as oracle and break the DBDH assumption with a non-negligible advantage.

**Proof** We will show that a simulator $\mathcal{B}$ can break the DBDH assumption with advantage $\frac{\epsilon}{2}$ if it takes an adversary $\mathcal{A}$, who can break our scheme in the Selective-Set model with advantage, as oracle.

The simulator $\mathcal{B}$ creates the following simulation

**Init:** The simulator $\mathcal{B}$ runs takes in a q-parallel BDHE challenge $\vec{y}$ ,T. $\mathcal{A}$ gives $\mathcal{B}$ a challenge ciphertext policy $W_1^*$.

**Setup:** To provide a public key PK to $\mathcal{A}$, $\mathcal{B}$ chooses random $\alpha' \in \mathbb{z}_p$ and implicitly sets $\alpha = \alpha' + a^{q+1}$ by letting $e(g,g)^\alpha = e(g^a, g^{a^q})e(g,g)^{\alpha'}$. For each x for $1 \le x \le U$ begin by choosing a random value $z_x$. Let X denote the set of indices i, such that $\rho^*(i) = x$. $\mathcal{B}$ run the program output $h_x$ as :

$$h_x = g^{z_x} \prod_{i \in X} g^{aM_{i,1}^*/b_i} \cdot g^{a^2 M_{i,2}^*/b_i} \cdots g^{a^n M_{i,n}^*/b_i}$$

Finally $\mathcal{B}$ sends $\mathcal{A}$ the public key.

**Phase 1:** $\mathcal{A}$ makes the following queries.

– **Extract( $L_1$ )[33]:** $\mathcal{A}$ submits an attribute list $L_1^* = (L_1, L_2, \cdots, L_n)$ in a secret key query. The attribute list must satisfy $L_1| \ne W_1^*$ or else $\mathcal{B}$ simply aborts and takes a random guess. $\mathcal{B}$ generates $SK_{L_1}^*$ and sends it to $\mathcal{A}$.

– **RKExtract $(L_2, W_2)$ :** $\mathcal{A}$ submits an attribute list $L_2^* = (L'_1, L'_2, \cdots, L'_n)$ and an access policy $W_2^*$ in a re-encryption key query. The attribute list must satisfy $L_2| \ne W_2^*$ or else $\mathcal{B}$ simply aborts and takes a random guess. In the same way, $\mathcal{B}$ generates $SK_{L_2}^*$ and sends it to $\mathcal{A}$.

Then $\mathcal{B}$ submits $L_1^*$ to **Extract** query and gets a secret key $SK_{L_1^*}$. Then it random choose $d^* \in \mathbb{z}_p$ and compute $g_2^{d^*}$, at last generate the $RK_{L_1 \to L_2}^*$ and sends it to $\mathcal{A}$.

**Challenge:** $\mathcal{A}$ submits two challenge messages $M_0$ and $M_1$. $\mathcal{B}$ flips a coin $b \in \{0, 1\}$. It creates

$$CT_{W_1} : C^* = M_b T \cdot e(g; g)^{s\alpha'};$$

$$C' = g^s; \quad g_2^s \ (C'_i, D'_i)_{1 \le i \le \ell}$$

**Phase 2**: **Phase 1** is repeated.

**Guess**: $\mathcal{A}$ outputs a guess $b'$ of b. $\mathcal{B}$ outputs 1 if and only if $b' = b$.

Therefore, the advantage of breaking the DBDH assumption is

$$Adv_\mathcal{A} = \left| \Pr[b' = b] - \frac{1}{2} \right|$$

$$= \left| \Pr[b = 0] \Pr[b' = b | b = 0] + \Pr[b = 1] \Pr[b' = b | b = 1] - \frac{1}{2} \right|$$

$$= \left| \frac{1}{2}\left(\frac{1}{2} + \epsilon\right) + \frac{1}{2}\frac{1}{2} - \frac{1}{2} \right| = \frac{1}{2}\epsilon$$

**Theorem 2** If there is an adversary who breaks our scheme in selective the reencryption key to get User1's $SK_{L_1}$ security model, a simulator can take the adversary as oracle and solve the DDH problem with a non-negligible advantage.

**Proof** We will show that a simulator $\mathcal{B}$ can solve the DDH problem with advantage $\epsilon$ if it takes an adversary $\mathcal{A}$, who can break our scheme in the selective the reencryption key security model with advantage, as oracle. Suppose the proxy colludes with the User2, he can get $g_2^{d^*}$ from User2.

Given a CBDH challenge tuple $[g, A, B, C] = [g, g^a, g^b, g^c]$ by the challenger, the simulator $\mathcal{B}$ creates the following simulation.

**Init**: The adversary $\mathcal{A}$ commits to a challenge attribute list $L_1^*$.

**Setup**: To provide a public key PK to $\mathcal{A}$, $\mathcal{B}$ generate

$$PK = g \ ; \ e(g; \ g)^{\alpha'}; \ g^{a'}; \ h_1, \ \cdots, \ h_u$$

**Queries**: $\mathcal{A}$ makes the following queries.

– Extract( $L_1^*$ ): $\mathcal{A}$ submits an attribute list $L'$ for a KeyGen query where $L_1^* \ne L'$, the challenger gives the adversary the secret key $SK_{L_1^*}$.

$$SK_{L_1^*}: K = g^{\alpha'} g^{a't'} \quad L = g^{t'} \quad \forall x \in L_1 \ K_x = h_x^{t'}$$

– RKExtract($L_2^*, W_2^*$): $\mathcal{A}$ submits an attribute list $L_2^*$ for a RKGen query, The challenger runs the **ReKeyGen** algorithm. Challenger choose randomly $d^* \in \mathbb{z}_p$ , computes $g_2^{d^*}$, then gives $g_2^{d^*}$, $RK_{L_1^* \to L_2^*}$ to $\mathcal{A}$.

$$RK_{L_1^* \to L_2^*}: K' = g^{\alpha'} g^{a't'} g_2^{a'd^*} \quad L' = g^{t'} g_2^{d^*}$$

$$\forall x \in L_1 \ K'_x = h_x^{t'} g_2^{d^*}$$

**Output**: $\mathcal{A}$ outputs the secret key $SK_{L_1^*}$ for the attribute list $L_1^*$, then $\mathcal{A}$ succeeds.

The advantage of $\mathcal{A}$ in this game is defined as

$$Adv_\mathcal{A} = \Pr[\mathcal{A} \text{ succeeds}].$$

If the proxy collude with the User2, he can get $g_2^{d^*}$ from User2, then he can get $K' = g^{\alpha'} g^{a't'} g_2^{a'd^*}$; $L = g^{t'}; \forall x \in L_1 \ K_x = h_x^{t'}$, $Adv_\mathcal{A}$ is as follows:

$$|\Pr[d^*, a' \xleftarrow{\mathcal{C}} \mathbb{z}_p : \mathcal{D}(g_2^{d^*}, g_2^{a'}, g_2^{a'd^*})] - \Pr[d^*, a', z \xleftarrow{\mathcal{C}} \mathbb{z}_p : \mathcal{D}(g_2^{d^*}, g_2^{a'}, g_2^{z})]|$$

## 4. Discussions

### 4.1 Re-Outsourcing Computation Reduction

In our scheme, the User1 can finish the process of proxy. If we omit the proxy, not only User1but also User2 would increase the amount of computation. We will use the third party. We can also outsource the policy of $W_2$.The algorithm is as follows.

**TransformKeyGenout(MSK; $L_2$)** The algorithm runs KeyGen(MSK; $L_2$) to obtain $SK_{L_2}$: $L'' = g^{t_1^{-1}} \forall x' \in L_2 \ K''_{x'} = h_{x'}^{t_1^{-1}}$. It chooses a random value $z \in \mathbb{z}_p^*$. It sets the transformation key TK as

$$L'' = g^{t_1^{-1}/z} \ \forall x' \in L_2 \ K''_{x'} = h_{x'}^{t_1^{-1}/z}.$$

and the private key SK as ($z$;TK).

After **Re-outsourcing ,** user2 get

$$(T_1 = e(g, g_2)^{\frac{-ads}{z}}, T_0 = \frac{e(g, g)^{\alpha s}}{e(g, g_2)^{-ads}})$$

Finally, User2 get $M = \frac{C}{T_0 \cdot T_1^z}$

### 4.2 Re-encryption Control

Note that if the User1 does not provide $g_2^s$ in ciphertext, the original decryption is not affected but the decryption of re-encrypted ciphertext cannot go on. That's because $g_2^s$ is only used in re-encrypted step, which is used in CP-policy. So she can control whether the ciphertext can be re-encrypted. In the same way, the proxy can also decide whether the re-encrypted ciphertext can be re-encrypted.

### 4.3 Extra Access Control

Our schemes try to allow the proxy to add extra access policy when re-encrypting ciphertext. For example, supposing the proxy can re-encrypt ciphertext under policy from $W_1$ to $W_2$, he can add an extra access policy$W_3$ to the re-encrypted ciphertext such that only user whose attribute list L simultaneously satisfies $W_2$ and$W_3$ can decrypt the re-encrypted ciphertext.

1. For a re-encryption key pair, choose a new $d' \xleftarrow{R} Zp$, compute new re-encryption key pair.

2. To decrease the amount of computation, add **Encrypt(PK, E($g_3^{d'}$), $W_3$)** to the re-encrypted ciphertext.

**Acknowledgements**

My deepest gratitude goes first and foremost to my supervisor, for her constant encouragement and guidance. She has walked me through all the stages of the writing of this thesis. Without her consistent and illuminating instruction, this thesis could not reach its present form.

Second my thanks would go to my beloved family for their loving considerations and great confidence in me all through these years. I also owe my sincere gratitude to my friends and my fellow classmates who gave me their help and time in listening to me and helping me work out my problems during the difficult course of the thesis.

**Keying Li** received a bachelor's degree from Linyi University, Shandong, China, in 2011. He is now pursuing Master of Faculty of science, Xidian University, Xi'an, China. His research interests cover the attributes based encryption, outsourcing, cloud computation, lossy trapdoor function, lossy encryption, and secret sharing.